\documentclass[%
 reprint,
superscriptaddress,
 amsmath,amssymb,
 aps,
pra,
]{revtex4-1}

\usepackage{graphicx}
\usepackage{dcolumn}
\usepackage{bm}
\usepackage{float}
\usepackage{mathrsfs}
\DeclareMathOperator\arctanh{arctanh}

\begin{document}

\preprint{APS/123-QED}

\title{Connection between inverse engineering and optimal control in shortcuts to adiabaticity}


\author{Qi Zhang}
\affiliation{International Center of Quantum Artificial Intelligence for Science and Technology (QuArtist) and Department of Physics, Shanghai University, 200444 Shanghai, China}
\affiliation{Laboratoire Collisions, Agr\'egats, R\'eactivit\'e, IRSAMC, Universit\'e de Toulouse, CNRS, UPS, France}

\author{Xi Chen}
\email{xchen@shu.edu.cn}
\affiliation{International Center of Quantum Artificial Intelligence for Science and Technology (QuArtist) and Department of Physics, Shanghai University, 200444 Shanghai, China}
\affiliation{Department of Physical Chemistry, University of the Basque Country UPV/EHU, Apartado 644, 48080 Bilbao, Spain}

\author{David Gu\'ery-Odelin}
\email{dgo@irsamc.ups-tlse.fr}
\affiliation{Laboratoire Collisions, Agr\'egats, R\'eactivit\'e, IRSAMC, Universit\'e de Toulouse, CNRS, UPS, France}

\begin{abstract}
We consider fast high-fidelity quantum control by using a shortcut to adiabaticity (STA) technique and optimal control theory (OCT). Three specific examples, including expansion of cold atoms from the harmonic trap, atomic transport by moving harmonic trap, and spin dynamics in the presence of dissipation, are explicitly detailed. Using OCT as a qualitative guide, we demonstrate how STA protocols designed from inverse engineering method, can approach with very high precision optimal solutions built about physical constraints, by a proper choice of the interpolation function and with a very reduced number of adjustable parameters.
\end{abstract}

\maketitle

\section{Introduction}
The last ten years witnessed the huge development of ``shortcuts to adiabaticity'' (STA) with wide applications ranging from atomic, molecular, and~optical physics (AMO) to quantum information transfer or processing~\cite{review1,review2}. The~concept of 
STA was originally proposed to speed up the adiabatic processes in quantum control. Nowadays, STA become versatile toolboxes for controlling the dynamics and transformation in quantum physics~\cite{review1,review2}, statistical  physics~\cite{DGO14,Mart1}, integrated optics~\cite{ShuoyenEPL}, and classical physics~\cite{Faure1,jarzynski1,jarzynski2,lianao}.
In this context, the~most popular STA techniques are the fast-forward scaling~\cite{masuda1,masuda2}, the~counterdiabatic driving~\cite{rice1,rice2} (or transitionless quantum algorithm~\cite{berry,chenprl2,adofolprl,deffner}), and~the invariant-based inverse engineering~\cite{chenprl}, and~their variants. These three techniques can be shown to be mathematically equivalent~\cite{Xipra11,Erikpra12}. However, the~diversity of the designs of shortcut protocols or their combination may be required for a realistic experimental implementation~\cite{multiple}. Furthermore, some counterdiabatic hamiltonians turn out to be unfeasible~\cite{3STIRAP}, or~some systems cannot be treated by means of invariant-based~engineering. 

STA method provides a useful toolbox for fast and robust quantum controls with applications in a wide variety of quantum platforms such as cold atoms~\cite{Schaff1,Schaff2}, NV~center spin~\cite{NV1,NV2} including for their use as a quantum sensor~\cite{jorge}, trapped ion~\cite{Kim}, and~superconducting qubit~\cite{Yingyi1,Yingyi2,Antti,Dapeng} to name a~few. 

Such controls have also a clear added value to quantum optimal control in quantum information processing and quantum computing~\cite{QuantumOC}, in~terms of analytical tools, numerical tools, and a combination of these two. Numerical optimal control such as the gradient ascent pulse engineering (GRAPE) algorithm works to some extent as a black box.
The dynamics and the structure of the control field are not easily predictable~\cite{Assemat}. STA techniques based on a clear physical picture deliver a more easily understandable framework but are mostly addressing problems of low complexity. However, these techniques have recently been combined with deep machine learning for more involved physical problems~\cite{HesonPNAS,Nature,Sels,Ding}.

Interestingly, shortcut protocols can be readily engineered to accommodate for various physical constraints {\cite{Larocca2020}}, or to mitigate an environmental noise. In this respect, the combination of inverse engineering methods and optimal control theory (OCT) has been particularly fruitful \cite{Stefanatostime-optimal,StepanatosTF,XiaojingPRA2014,Xipra11optimal,Qipra,QiJPB,PRL2013D,njp2012,inverse13}. 
Most STA techniques provides solutions that are robust against a small variation of the duration of the parameter engineering. In Ref.~\cite{Martikyan}, it is shown how OCT solutions can be adapted to accommodate for extra boundary conditions to ensure a similar robustness. Alternatively, the STA technique of Ref.~\cite{DavidMuga} provides an explicit solution for linear control problems fulfilling the Kalman criterium \cite{Martikyan2}.

In this article, we compare systematically the inverse engineering method with the result of optimal control theory on three specific examples that can be addressed analytically in both formalisms: expansion of cold atoms from the harmonic trap, atomic transport by moving harmonic trap, and spin dynamics in the presence of dissipation. Our aim is to provide a pedagogical introduction and comparison between a simple if not the simplest Shortcut To Adiabaticity technique, the direct inverse engineering of the equation of motion of the dynamical variables, and the optimal control theory. STA techniques are built about the boundary conditions while OCT involves the minimization of a cost function. To facilitate the comparison we therefore discuss how inverse engineered (IE) solutions can be modified in order to minimize a cost function and mimic OCT solutions. Similarly to the variational method in quantum mechanics, and as illustrated in the following, the family of functions over which the minimization is performed play a crucial role. In the following, we also show how a simple ansatz having just a few tunable parameters can approach very precisely the optimal solution obtained for a given physical constraint.


\section{Fast cooling in time-varying harmonic traps}

Fast frictionless coooling for ultracold and Bose-Einstein condensates belongs to the first experimental demonstrations of STA techniques \cite{Schaff1,Schaff2}. Such techniques have been subsequently adapted and applied to cold-atom mixtures \cite{Choi}, Tonks-Girardeau gas \cite{CampoPRA11,deffner}, Fermi gases \cite{Deng,DGO14}, and many-body systems \cite{Jorge15}. 

In this section, we address the problem of fast atomic cooling in a time-dependent harmonic trap \cite{chenprl}. We derive the time-dependence of the trap frequency by an inverse engineer procedure on an Ermakov equation and using OCT. We subsequently compare the two types of solutions. Interestingly, the tunability inherent to the inverse engineering method provides the required flexibility to shape the inverse-engineered trajectories to minimize a cost function. We show how such solutions can be simply adapted to get results very close to optimal solutions for a time-averaged energy cost function \cite{XiMugaPRA2010,chenprl,Stefanatostime-optimal}.

\subsection{Model, Hamiltonian, and the Inverse Engineering Approach}

\label{STAharmonicCooling}

More specifically, we consider in the following the fast decompression of a one-dimensional (1D) harmonic potential from an initial angular frequency $\omega(0)=\omega_0$ to the final target one $\omega(t_f)=\omega_f$, ($\omega_f<\omega_0$). The problem amounts to finding the time-dependent solution of the Schr\"odinger equation that ensures the transformation from the ground state of the initial trap to the ground state of the final trap in a finite amount of time $t_f$: 
\begin{equation}
\label{Schor1}
i \hbar\frac{\partial\psi}{\partial t}=\left[-\frac{\hbar^2}{2m}\frac{\partial^2}{\partial x^2}+\frac{1}{2}m \omega^2(t) x^2\right]\psi.
\end{equation}
For this purpose, we look for a scaling solution of the form $\psi(x, t)=\exp[-\beta(t)] \exp[-\alpha(t) x^2]f(\rho(t)=x/b(t), t)$. The first factor accounts for the normalization, the second factor for the evolution of the phase (we show below that it is purely imaginary) and the last one for the desired scaling dynamics. By plugging such an ansatz into the Schr\"odinger equation, we find how the different parameters are related:
\begin{eqnarray}
\nonumber
i \hbar\partial_t f&=&\left(i \hbar \dot \beta+\frac{\hbar^2}{m}\alpha \right)f+\left(\frac{2\hbar^2}{m}\alpha+i\hbar\frac{\dot b}{b}\right)\rho \partial_{\rho}f
\\
&+& \!\! \left( \! i\hbar\dot \alpha \!-\! \frac{2\hbar^2}{m}\!+\! \frac{1}{2}m\omega^2 \! \right) \! b^2 \rho^2 f-\frac{\hbar^2}{2m b^2}\partial_{\rho \rho}f.
\end{eqnarray}
By introducing the renormalized time $\tilde{t}(t)=\int^t_0 dt'/b(t')^2$ and for the choice $\alpha=(-i m/2\hbar)\dot b/b$ and $\beta=\ln b/2$, the effective wave function $\Psi(\rho, \tilde{t})=f(\rho, t)$ obeys a \emph{time-independent}  Schr\"odinger equation:
\begin{equation}
\label{Schro2}
i \hbar\frac{\partial\Psi}{\partial \tilde{t}}=\left[-\frac{\hbar^2}{2m}\frac{\partial^2}{\partial \rho^2}+\frac{1}{2}m \omega^2_0 \rho^2 \right]\Psi,
\end{equation}
provided that the scaling parameter $b(t)$ satisfies the following Ermakov equation
\begin{equation}
\label{Ermakov}
\ddot b+\omega^2(t)b=\frac{\omega^2_0}{b^3}.
\end{equation}
{Interestingly, this latter equation is amenable to a set of linear equations. Indeed, it is the equation of an effective 2D oscillator in polar coordinates, the $1/b^{-3}$ is nothing but the centrifugal barrier which acts as a repulsive force that prohibits the access to a zero value of $b$.} Alternatively, the very same result can be obtained by using Lewis-Riesenfeld dynamical invariant \cite{chenprl}.  The ground state wave function in such a time-dependent harmonic trap reads
\begin{equation}
\psi(x, t)=\frac{\mathscr{N}}{\sqrt b}\exp\left(\frac{i m\dot b}{2\hbar b} x^2\right) \exp\left(-\frac{x^2}{2 a^2_0 b^2}\right),
\label{eqgs}
\end{equation}
where $\mathscr{N}$ accounts for the normalization and $a_0=\sqrt{\hbar/(m\omega_0)}$. The self-consistent boundary  conditions for a smooth continuous interpolation function are \cite{chenprl}:
\begin{eqnarray}
\nonumber
b(0) &=& 1,\;\;\dot b(0)=0,\;\;\ddot b(0)=0,\\
\label{bInitial}
b(t_f) &=& \gamma=\sqrt{\omega_0/\omega_f},\;\;\dot{b}(t_f)=0,\;\;\mbox{and}\;\; \ddot b(t_f)=0.
\end{eqnarray}
As a simple example, one can choose for the scaling factor $b(t)$ a fifth order polynomial ansatz that fulfills the above six boundary conditions \cite{chenprl}: 
\begin{equation}
\label{expanhar-xcpoly1}
b(\tau)=1+(\gamma-1) (10 \tau^3 - 15 \tau^4 + 6 \tau^5).
\end{equation}

 In view of the comparison with optimal protocols, we calculate hereafter the mean energy associated to the ground state wave function (\ref{eqgs}) \cite{XiMugaPRA2010}:
\begin{eqnarray}
\nonumber
\overline{E} &\equiv & \frac{1}{t_f}\int_0^{t_f}E(t) dt=\frac{1}{t_f}\int_0^{t_f} \langle \psi (t)|H(t)| \psi(t)\rangle dt
\\
\label{Expansionhar-En}
&=& \frac{\hbar}{2\omega_0}\frac{1}{t_f} \int_0^{t_f} \left(\dot{b}^2+ \frac{\omega_0^2}{b^2}\right) dt,
\end{eqnarray}
where $E(t) = K(t) + E_p(t)$  is the sum of the kinetic energy  $K(t)= \hbar(\dot{b}^2+ \omega_0^2/b^2)/(4\omega_0) $ and the potential energy $E_p=\hbar \omega^2(t) b^2/(4\omega_0)$. The mean energies obey the Virial theorem: $\overline{E_p}=\overline{K}=\overline{E}/2$. For any $b(t)$ trajectory that fulfills the boundary conditions, one can infer $\omega(t)$ from Eq.~(\ref{Ermakov}) and calculate explicitly the mean energies. Using the available freedom to shape the scaling factor $b(t)$, the inverse-engineered solutions can be tuned so to minimize the time-averaged energy as discussed in section \ref{comparison}.

\subsection{Optimal control theory}

Shortcut To Adiabaticity protocols such as  inverse engineering are built about the boundary conditions. We have provided an example using a polynomial interpolation. Optimal control theory (OCT) offers an alternative to find a path between two states but shall be built about a cost function. We propose hereafter to use OCT on the Ermakov equation (\ref{Ermakov}). 

For this purpose, we recast Eq.~(\ref{Ermakov}) into a set of first order nonlinear coupled equations, $\dot{\textbf{x}} = \textbf{f}(\textbf{x}(t), u)$ by defining the $\textbf{x}$ components as  $x_1 = b(t)$ and $x_2 = \dot{b}/\omega_0$, and introducing the (scalar) control function,
$u(t) = \omega^2(t)/\omega_0^2$:
\begin{eqnarray}
\label{system-1}
\dot{x}_1  &=&  x_2,
\\
\label{system-2}
\dot{x}_2 &=& - u x_1+\frac{1}{x_1^3}.
\end{eqnarray}

In the following, we work out two OCT solutions associated to the minimization of the final time and then of the mean energy. {As a result of the nonlinear character of the set of Hamiltonian equations, the Pontryagin maximum principle only gives a necessary condition to get an extremum.}

\subsubsection{Time-optimal solution}

The so-called time-optimal solution amounts to minimizing the cost function
\begin{equation}
\label{cost}
J = \int^{t_f}_0 1 dt.
\end{equation}
with the boundary conditions (\ref{bInitial}) which translates on the $\textbf{x}$ vector components as $x_1(0)=1$, $x_2(0)=0$ and $x_1(t_f)=\gamma$ and $x_2(t_f)=0$. We furthermore choose the constraint $ |u| \leq 1 $ \cite{Hoffman,Stefanatostime-optimal}. {We stress that we let the possibility for the control parameter to be either positive or negative. When it is negative, the curvature of the harmonic confinement is reversed. Atoms are therefore transiently expelled which provides a method to accelerate the desired transformation.}

\begin{figure}[t]
	\centering
	\centering\includegraphics[width=0.45\textwidth]{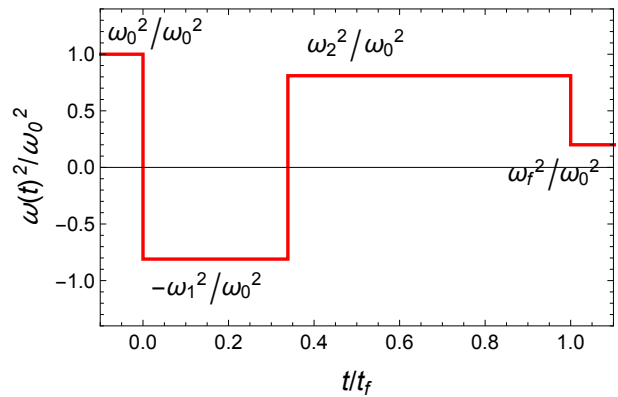}
	\caption{Fast cooling in time-varying harmonic traps: The 3-jump ``bang-bang" control function, $u(t)=\omega^2(t)/\omega_0^2$.}
	\label{fig1}
\end{figure}

To minimize the cost function (\ref{cost}), we apply the Pontryagin maximum principle which states that there exists non-zero, continuous vector $\textbf{p}$ with components ($p_0, p_1, p_2$), fulfilling Hamilton's equations \cite{Stefanatostime-optimal,StepanatosTF}:
$ \dot{\textbf{x}} = \partial H_c/\partial \textbf{p}$ and $\dot{\textbf{p}} = - \partial H_c/\partial \textbf{x}$.
With the cost function $J$, the control Hamiltonian $H_c$ reads

\begin{equation}
\label{controlH}
H_c = p_0 + p_1 x_2 + p_2 \left( -x_1 u+\frac{1}{x_1^3}\right),
\end{equation}
where $p_0$ is a non-zero normalization constant, and $p_1$ and $p_2$ are generalized Lagrange multipliers.
The Pontryagin's maximum principle states that at any instant ($0 \leq t \leq t_f$), the values of the control function $u$ maximize $H_c$. As $H_c$ is linear in the control function $u$ and since $x_1>0$, the sign of the factor in front of $u$, ($-p_2 x_1$) is fully determined by the sign of $-p_2$. This latter parameter plays the role of a switching function for ``bang-bang" type control as discussed in the literature \cite{XiaojingPRA2014,Hoffman,Stefanatostime-optimal,StepanatosTF}. {The fact that the Hamilton equations are nonlinear enables the possibility to have multiple bang-bang solutions \cite{Stefanatostime-optimal}. We consider in the following the simplest solution with analytical expression. This ``bang-bang"  solution has a single intermediate time (see Fig.~\ref{fig1}):}

\begin{figure}[t]
	\centering
	\centering\includegraphics[width=0.5\textwidth]{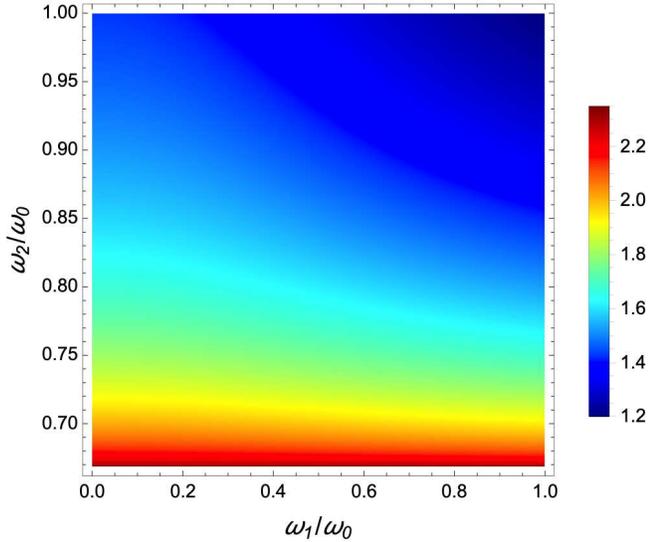}
	\caption{Fast cooling in time-varying harmonic traps: 2D color plot of the final normalized time $ t_f \omega_0 $ for a 3-jump ``bang-bang" control as a function of the first pulse amplitude $\omega_1/\omega_0$, and the second pulse amplitude $\omega_2/\omega_0$.}
	\label{fig3}
\end{figure}

\begin{equation}
\label{u3-jump}
u(t)= \left\{
\begin{aligned}
1 &,& ~ t\leq 0 \\
-(\omega_1/\omega_0)^2 &,& ~ 0< t<t_1 \\
(\omega_2/\omega_0)^2 &,& ~ t_1< t<t_f \\
(\omega_f/\omega_0)^2 &,& ~ t\geq t_f
\end{aligned}
\right.
\end{equation}
With such a control function, we infer the value of the scaling factor $b(t)$ from the Ermakov equation 
and  find the following solution for ``bang-bang" control that fulfills the boundary conditions (\ref{bInitial}):

\begin{eqnarray}
\label{3jump-b}
b(t) \!=\! \left\{\begin{array}{ll}
\sqrt{1+\frac{\omega_1^2+\omega_0^2}{\omega_1^2} \sinh^2(\omega_1 t)}, & 0 \! \leq \! t \! \leq \! t_1
\\
\sqrt{\gamma^2+\frac{\omega_0^2-\gamma^4\omega_2^2}{\gamma \omega_2^2} \sin^2[\omega_2 (t_f-t)]}, & t_1 \! \leq \! t \! \leq \! t_f.
\end{array}\right.
\end{eqnarray}
It is worth noticing that the Ermakov equation implies that the quantity $x_2^2+ux_1^2+x_1^{-2}=c$ is constant. The value of the constant $c$ is fixed by the initial conditions  for $0< t<t_1$ and by the final conditions for $t_1< t<t_f$. Using the continuity of $b(t)$ at $t_1$ and $t_f$ due to the second derivative in the Ermakov equation, we find the explicit expression for both times \cite{Hoffman,Stefanatostime-optimal}:

\begin{eqnarray}
\label{t1}
t_1 &=& \frac{1}{\omega_1}{\rm{arcsinh}} \sqrt{\frac{\omega_1^2(\gamma^2-1) (\gamma^2\omega_2^2-\omega_0^2)}{\gamma^2(\omega_1^2+\omega_0^2) (\omega_2^2+\omega_1^2)}}, \\
\label{tf}
t_f &=& t_1+\frac{1}{\omega_2} \arcsin{\sqrt{\frac{\omega_2^2(\gamma^2-1)(\gamma^2 \omega_1^2+\omega_0^2)}{(\omega_1^2+\omega_2^2) (\gamma^4\omega_2^2-\omega_0^2)}}}.
\end{eqnarray}
As the time $t_1$ shall remain real, we deduce from Eq.~(\ref{t1}) that $\omega_2 \geq \omega_0/\gamma>\omega_f$. The last inequality is naturally satisfied because of the cooling constraint $\omega_0>\omega_f$. The first inequality requires $\omega_0/\gamma \leq \omega_2 \leq \omega_0$.
In Fig. \ref{fig3}, we plot the normalized final time $s_f= t_f\omega_0 $ as a function of $\omega_1/\omega_0$ and $\omega_2/\omega_0$ in their accessible domains. We conclude that the shortest normalized final time $s_f$ is obtained for the largest $\omega_1$ and $\omega_2$. With the choice $\omega_2=\omega_1=\omega_0$, we obtain the shortest time

\begin{equation}
s_f^{min}=\frac{\pi}{4}+\frac{1}{2}\ln \left(\frac{\omega_0}{\omega_f}\right).
\end{equation}
The lowest bound for $\omega_2$, namely, $\omega_2=\omega_0/\gamma$
provides the upper bound for final time

\begin{equation}
s_f=\frac{\pi}{2}\gamma,
\end{equation}
where the first period of time is reduced to $t_1=0$, so that only two jumps are needed. 

\begin{figure}[t]
	\begin{center}
	\centering\includegraphics[width=0.4\textwidth]{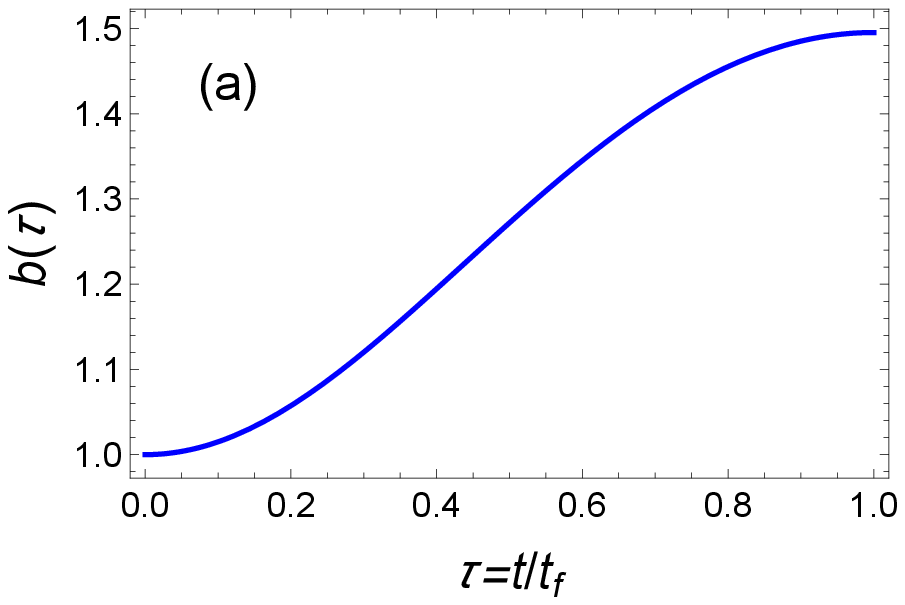}
		\centering\includegraphics[width=0.4\textwidth]{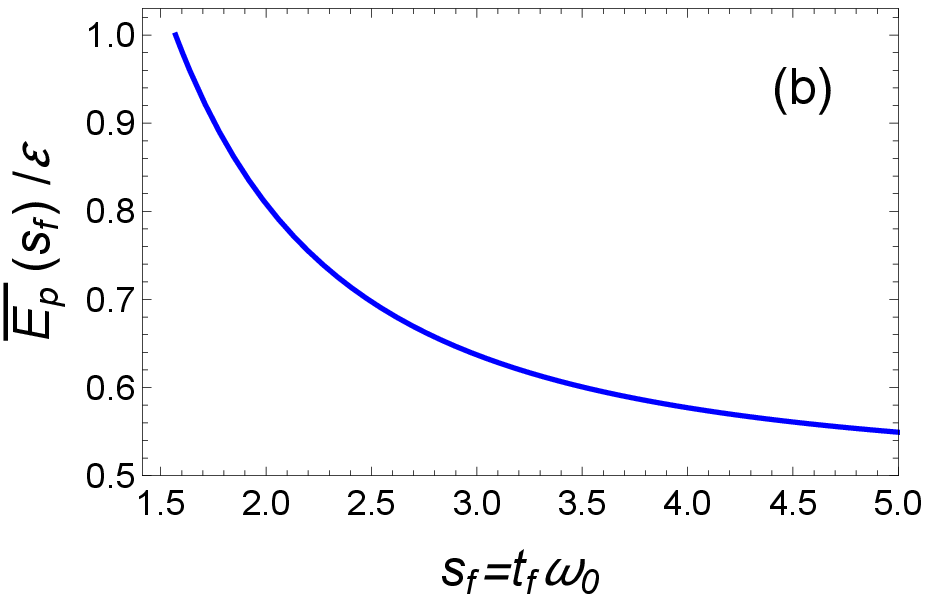}
		\caption{\label{fig1bE} Fast cooling in time-varying harmonic traps: (a)Example of time-optimal trajectory of $b(t)$ from Eq. (\ref{3jump-b}). Parameters: $\omega_f^2=\omega_0^2/5$. (b) The time-averaged energy as a function of the normalized final time $s_f=\pi\gamma/2$, obtained from the time-optimal control solution.}
	\end{center}
\end{figure}

In this latter range of parameter, the scaling factor reads

\begin{equation}
\label{ExpanharBB-b}
b(\tau) = \sqrt{\gamma^2+(1-\gamma^2)\sin^2\left[\frac{\pi(1-\tau)}{2}\right]},
\end{equation}
with $\tau=t/t_f$. In Fig.~\ref{fig1bE} (a), we plot such an example of the time evolution of $b(t)$. The solution that corresponds to the upper bound for the final time also provides the minimum time-averaged energy. Using Eq.~(\ref{Expansionhar-En}), we calculate this latter quantity:

\begin{equation}
\overline{E_p}= \frac{\varepsilon}{2}\left(1+ \frac{1}{\gamma^2}\right)=  \frac{\varepsilon}{2}\left(1+\frac{\pi^2}{4 s_f^2}\right),
\end{equation}
where $\varepsilon=\hbar\omega_0/4$. In Fig.~\ref{fig1bE} (b), we plot this time-averaged energy $\overline{E_p}$ as a function of the final time $s_f$. It is worth noticing that $\omega_f$ and $t_f$ are not independent since $s_f=\pi\gamma/2=\pi \sqrt{\omega_0/\omega_f}/2$.

\subsubsection{Time-averaged energy minimization}

\begin{figure}[t]
	\centering\includegraphics[width=0.45\textwidth]{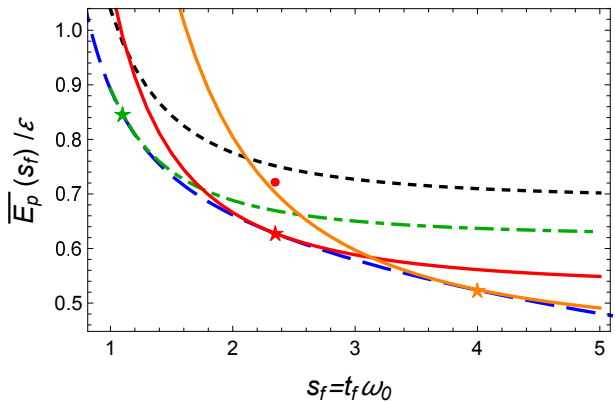}
	\caption{Fast cooling in time-varying harmonic traps: Comparison of time-averaged potential energy for different optimal protocols: (1) energy-minimization (blue dashed line), (2) time-optimal protocol (with $\omega_f$ fixed, $\overline{E_p}$ constant) (red point); and  inversed-engineered protocols: (1) 0-freedom polynomial in Eq.~ (\ref{expanhar-xcpoly1}) (black dotted line), (2) polynomial IE solution with two free parameters optimized for a given normalized final time (stars) (see Table \ref{tablesf}): $s_f=1.1$ (green line), $s_f=\pi \gamma/2$ (red solid line), and $s_f=4$ (orange solid line). Parameters: $\omega_f^2=\omega_0^2/5$.}
	\label{fig2E}
\end{figure}

In this section, we consider optimal control solution associated to the minimization of time-averaged energy with unbounded constraint \cite{XiMugaPRA2010}. The lower bound for the time-averaged potential (total) energy in Eq.~(\ref{Expansionhar-En}) reads \cite{XiMugaPRA2010,Cui}:

\begin{eqnarray}
\nonumber
\overline{E_p}^{op} &= & \varepsilon
\left[ \left(\frac{B}{s_f}\right)^2-1-\frac{2}{s_f} \arctanh \left(\frac{B^2+B-s_f^2}{s_f}\right) \right.
\\
\label{ExpanHarOCTu-Ep}
&+& \left. \frac{2}{s_f} \arctanh \left(\frac{B}{s_f}\right) \right],
\end{eqnarray}
with the following solution of $b(\tau)=\sqrt{(B^2-s_f^2)\tau^2+2B\tau+1}$ and $B=-1+\sqrt{s_f^2+\gamma^2}$. In Fig. \ref{fig2E}, we plot this lower bound for optimized time-averaged energy as a blue dashed line.

\subsection{Comparison between IE and OCT}
\label{comparison}
In the previous subsections, we have reviewed the streamline of IE and OCT protocols to ensure a fast frictionless decompression in a harmonic trap whose strength can be time-engineered. As already discussed, there is a lot of freedom to design inverse-engineered protocols since the only requirements concern the boundary conditions. However, the question of the mean energy cost of such protocols may be relevant since a real potential always exhibits some anharmonicity when the potential energy becomes too large. In what follows, we propose to design IE protocols having a minimal mean potential energy. We will show how we can readily approach the optimal results. 

The IE solution exhibited in Eq.~(\ref{expanhar-xcpoly1}) relies on a fifth-order polynomial that fulfills the six boundary conditions. In Fig.~\ref{fig2E},  we plot the corresponding mean potential energy $\overline{E_p}(s_f)$ using a black dotted line which turns out to be quite far from the optimal solution (dashed blue line).

To reduce $\overline{E_p}(s_f)$, we remove the constraints on $\dot{b}$ and $\ddot{b}$ at initial and final time since they are not strictly speaking necessary neither fulfilled by the optimal solution. We also enlarge the parameter space for $b(\tau)$ using a third-order polynomial ansatz 
{$b(\tau)=\sum\nolimits_{n=0}^{3} a_n \tau^n$} to keep some free parameters. The two boundary conditions yields $a_0=1$ and $a_1=-1-a_2-a_3+\gamma$. For different normalized final time $s_f$, we can therefore minimize the time-averaged energy with respect to the two parameters $a_2$ and $a_3$. In Table \ref{tablesf}, we provide the optimal values $a_2$ and $a_3$ that minimizes the mean potential energy for the three cases with $s_f=1.1$, $s_f=\pi \gamma/2$, and $s_f=4$. 

\begin{figure}[t]
	\centering\includegraphics[width=0.35\textwidth]{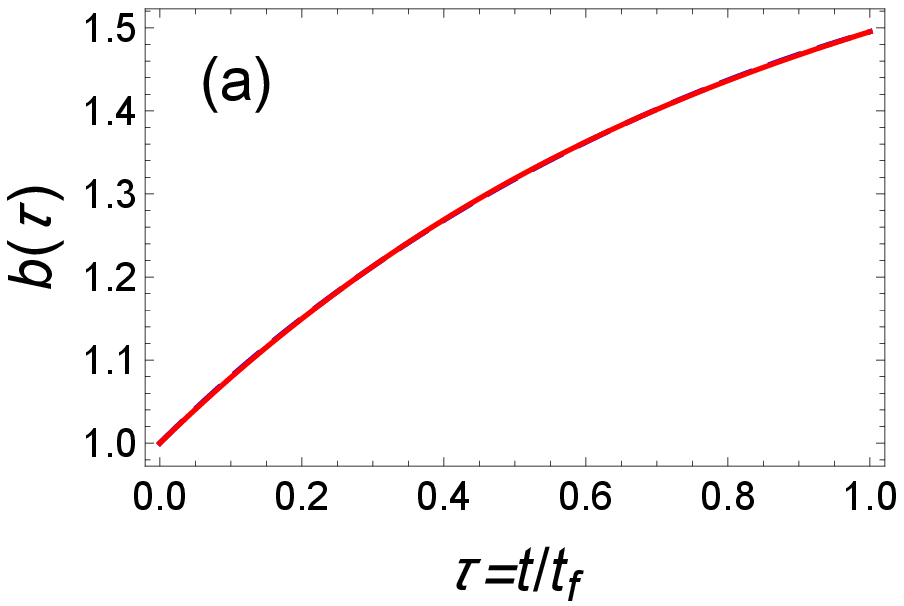}
	\centering\includegraphics[width=0.35\textwidth]{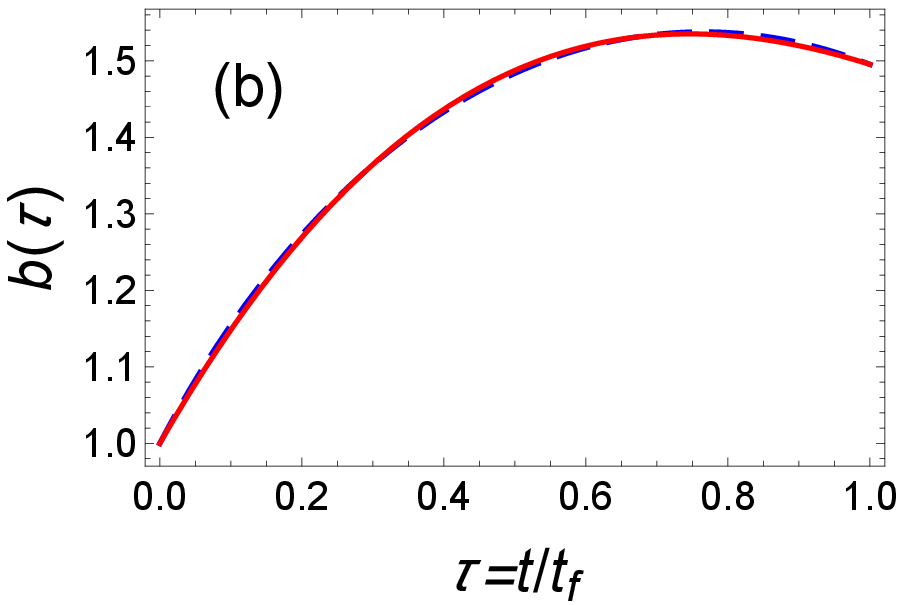}
	\centering\includegraphics[width=0.35\textwidth]{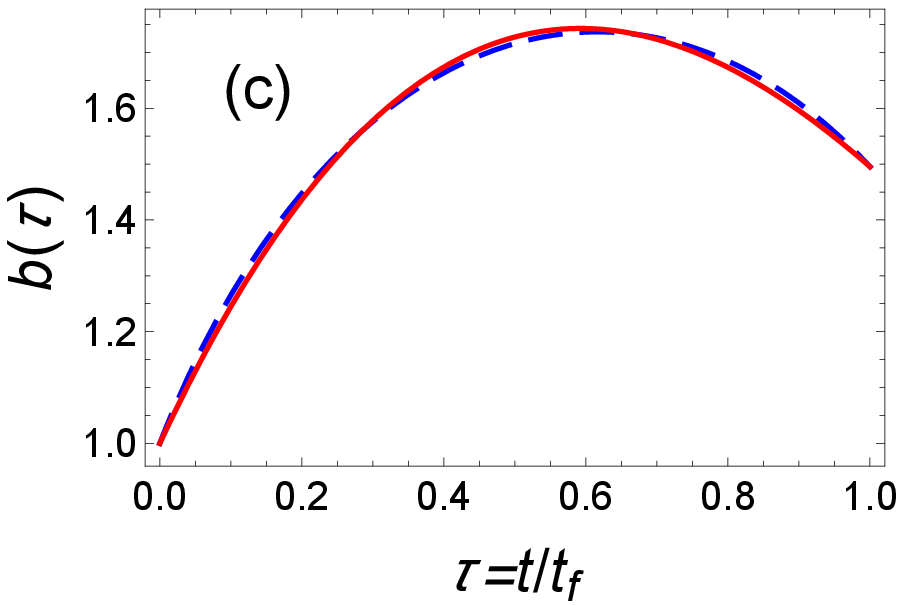}
	\caption{Fast cooling in time-varying harmonic traps: Comparison of time-dependent normalized variable $b(\tau)$ obtained from optimal control theory (averaged energy optimization) (blue dashed line) and from inverse engineered solutions optimized to minimize the time-averaged energy (red solid line) for different final times (a) $s_f=1.1$, (b) $s_f=\pi \gamma/2$, and (c) $s_f=4$. The corresponding optimal values of polynomial functions are detailed in Table \ref{tablesf}. Parameters: $\omega_f^2=\omega_0^2/5$.}
	\label{fig3b}
\end{figure}

\begin{table}[H]
	\centering
	\setlength{\tabcolsep}{7mm}{
	\begin{tabular}{ccc}
		\hline
		\textbf{$s_f$} & \textbf{$a_2$} & $a_3$ \\
		\hline
		1.1 & -0.44893 & 0.10996 \\
		$\pi\gamma/2$ & -1.47741 & 0.34535 \\
		4 & -2.86194 & 0.62841 \\
		\hline
	\end{tabular}}
	\caption{\label{tablesf} Optimal values of the free parameters $a_2$ and $a_3$ in the three-order polynomial ansatz for the IE protocol that minimize the time-averaged energy. Parameter $\omega_f^2 = \omega_0^2/5$.}
\end{table}

The results are represented as stars in Fig.~\ref{fig2E}. They nearly coincide with the result of the optimal control theory. This is confirmed by plotting the scaling functions for both protocols (see Fig.~\ref{fig3b}). We conclude that the IE trajectories inspired by the OCT solutions can be readily designed to approach with an impressive accuracy the exact OCT solutions. 


\section{Fast transport of atoms in moving harmonic traps }

STA techniques have also been applied to high-fidelity fast quantum transport of neutral atoms \cite{DGO08} or charged ions \cite{Walther12,Bowler12} using a moving trap. Such developments have a wide range of applications from quantum information processing \cite{Erikion,ErikBEC} to atom fountain clock, atom chip manipulation \cite{CorgierNJP,Becker,Amri} or atomic interferometry \cite{Dupont}. In recent closely related works, optimal trajectories that minimize the excitation in ion shuttling in the presence of stochastic noise have been designed by combining invariant-based inverse engineering, perturbation theory, and optimal control \cite{Xiaojingpra,Xiaopra18}. 

In this section, we address the problem of the fast transport of a single atom based on a moving 1D harmonic potential. The particle is supposed to be initially in the ground state and shall remain in the ground state at the final time. We follow the same kind of presentation as previously: We first design inverse-engineered  protocols, we then derive the OCT protocols for time  \cite{Xipra11optimal} and mean-energy optimization \cite{QiJPB}, and eventually compare both approaches.

\subsection{Classical and quantum inverse-engineered solutions}
The time-dependent Hamiltonian of atomic transport using a moving harmonic trap reads

\begin{equation}
\label{Hamiltonian}
H (t)= \frac{\hat{p}^2}{2 m} + \frac{1}{2}m \omega^2_0 [\hat{x}-x_0(t)]^2,
\end{equation}
where 
$\omega_0$ is the constant trap angular frequency,
and $x_0(t)$ the time-dependent position of the trap center. This problem amounts to finding the appropriate driving of this harmonic oscillator. The exact mapping between the classical and quantum solutions enables one to solve the classical problem to get a solution valid quantum mechanically \cite{DavidMuga}. The time-evolution of the coordinate, $x(t)$, of a classical particle under the time-dependent Hamiltonian (\ref{Hamiltonian}) is given by

\begin{equation}
\label{differential}
\ddot{x}+\omega_0^2(x-x_0(t))=0.
\end{equation}
{A smooth perfect transport i.e. a transport without any residual oscillations at final can be obtained using inverse engineering by imposing the six boundary conditions: }
\begin{eqnarray}
\nonumber
x(0) = x_0(0)=0,\;\;\dot{x}(0)=0,\;\;  \ddot{x}(0)=0,\;\;
\\
\label{con0}
x(t_f) \!=\! x_0(t_f)\!=\!d,\;\; \dot{x}_0(t_f)\!=\!0,\;\; \mbox{and}\;\; \ddot{x}_0(t_f)\!=\!0.
\end{eqnarray}
Any interpolation function $x(t)$ that fulfills these boundary conditions provides a possible solution of our problem. For instance, one can use the following fifth order polynomial interpolation function:
\begin{equation}
x(t)=d \left[   10 \left(\frac{t}{t_f}\right)^3 -15\left(\frac{t}{t_f}\right)^4+6\left(\frac{t}{t_f}\right)^5   \right].
\label{5orderpoly}
\end{equation}
Once $x(t)$ is known, the trajectory of the trap center $x_0(t)$ can be directly inferred from Eq.~(\ref{differential}). A similar result can be derived quantum mechanically using the properties of dynamical invariants \cite{Erikion,Xipra11optimal}. In view of the optimization that we will perform later on, it is worth working out the instantaneous average potential energy
\begin{equation}
\label{potential}
\langle V(t)\rangle=\frac{\hbar\omega_0}{4}+E_p(t),
\end{equation}
where the first term accounts for the zero-point energy contribution and $E_p(t)= \frac{1}{2} m\omega_0^2(x(t)-x_0(t))^2$ i.e. the instantaneous potential energy for the effective classical particle. The time-averaged potential energy is defined by 

\begin{equation}
\label{Ep}
\overline{E_p}= \frac{1}{t_f} \int^{t_f}_0 E_p(t) dt.
\end{equation}

\subsection{Optimal control theory}
To recast this problem as an optimal problem, we  define the variables $x_1 (t)= x(t)$ and $x_2(t) = \dot{x}$, and the control function $u(t) = x(t) -x_0(t)$. The control function corresponds to the relative position of the effective particle with respect to the trap center. The equation of motion (\ref{differential}) for the effective particle can be encapsulated in the following set of linearly coupled first order differential equations $\dot{\textbf{x}} = \textbf{f} [\textbf{x}(t),u]$:
\begin{eqnarray}
\label{system-10}
\dot{x}_1  &=&  x_2,
\\
\label{system-20}
\dot{x}_2 &=& - \omega_0^2 u.
\end{eqnarray}
{Interestingly, for  this linear system, the solution deduced from the Pontryagin formalism provides the unique control solution $u$ that minimizes the cost function.}

\subsubsection{Time minimization}

In this section, we solve the time-optimal problem with an upper bound on the relative displacement $|u|\leq \delta$. The cost function to minimize $t_f$ is

\begin{equation}
J = \int^{t_f}_0 1 dt.
\end{equation}
The corresponding Pontryagin Hamiltonian reads $
H_c = p_0+p_1 x_2-\omega^2 p_2 u$, 
where the Lagrange multipliers $p_1$ and $p_2$  fulfill $\dot{p}_1 = 0$ and $\dot{p}_2 =- p_1$. We deduce $p_1=c_1$ and $p_2=-c_1 t+c_2$ where $c_1$ and $c_2$ are constants to be determined. The Hamiltonian $H_c$ is a linear function of the bounded control function $u(t)$. As a result,  the sign of $p_2$ sets the sign of $u(t)$ to maximize $H_c$. 
The parameter $p_2$ being a linear function of time, the sign of $p_2$ can only change once. By considering the initial and final boundary conditions, the appropriate control sequence taking into account the upper bound for $|u(t)|$ is a (three-jump) ``bang-bang" control
\begin{eqnarray}
\label{control function-bangbang}
u (t) = \left\{\begin{array}{lll}
0, & t \leq 0\\
-\delta, & 0<t<t_1\\
\delta, & t_1<t<t_f\\
0, & t \geq t_f .
\end{array}\right. 
\end{eqnarray}
With such a control function, the time-optimal solution of Eq.~(\ref{differential}) compatible with the boundary conditions (\ref{con0}) reads

\begin{eqnarray}
\label{xc_time-bang}
x (t) = \left\{\begin{array}{lll}
0, & t \leq 0
\\
\omega^2 \delta t^2/2, & 0<t<t_1
\\
d-\omega^2 \delta (t-t_f)^2/2, & t_1<t<t_f\\
d. & t \geq t_f.
\end{array}\right.
\end{eqnarray}
The driving of the trap bottom is then given by $x_0(t)=\ddot{x}(t)/\omega^2_0+x(t)$. By imposing, the continuity on $x(t_1)$ and $\dot{x}(t_1)$, one gets the explicit expression for the switching and final times:

\begin{equation}
t_1=\frac{t_f}{2},~~~
t_f=\frac{2}{\omega_0}\sqrt{\frac{d}{\delta}}.
\end{equation}
According to Eq.~(\ref{Ep}), the time-averaged potential energy $\overline{E_p}$ for this constrained protocol is

\begin{equation}
\label{EnTime}
\overline{E_p}=\frac{8 m d^2}{\omega_0^2 {t_f}^4}=\frac{1}{2}m \omega_0^2\delta^2.
\end{equation}

\subsubsection{Mean potential energy minimization}

In this section, we work out the energy-optimal protocol.  We here provide a solution that minimizes the time-averaged potential energy for a given transport time $t_f$ and distance $d$, with unbounded constraint. According to the definition of potential energy, $E_p=\frac{1}{2}m \omega^2_0(x-x_0)^2$, the cost function for this problem is

\begin{equation}
J = \frac{1}{2}m \omega^2_0\int^{t_f}_0 u^2 dt,
\end{equation}
and the Pontryagin Hamiltonian 

\begin{equation}
H_c = \frac{1}{2}m \omega^2_0 p_0 u^2 + p_1 x_2 - p_2 \omega^2_0 u.
\end{equation}
The Hamilton equations give two costate equations similar to those derived in the previous section. For the normalization, we can choose the constant parameter $p_0=-1/m$, so that the optimal problem amounts to maximizing the quantity $-u^2/2-p_2 u$.

For convenience, we consider the unbounded case  ($u(t)$ is unbounded) which sets the lowest bound for time-averaged potential energy $\overline{E_p}$. The quantity $-u^2/2-p_2 u$ is maximal for $u=-p_2$. This expression for the control fonction combined to Eq.~(\ref{differential}) and the boundary conditions (\ref{con0}) enables one to determine the optimal trajectory of the center of mass:

\begin{equation}
\label{xc-unbounded}
x (t) =  \frac{d t^2 }{{t_f}^2} \left(3-2 \frac{t}{t_f}\right),
\end{equation}
from which we infer the trap center trajectory $x_0(t)$ using Eq.~(\ref{differential}) with initial and final boundary conditions $x_0(0)=0$ and $x_0(t_f)=d$:

\begin{figure}[t]
	\begin{center}
		\centering\includegraphics[width=0.4\textwidth]{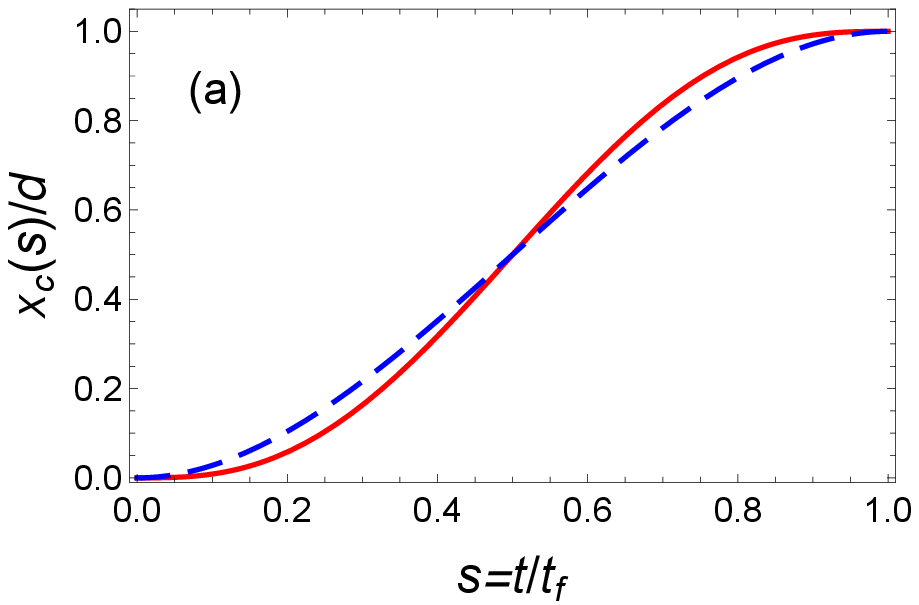}
		\centering\includegraphics[width=0.4\textwidth]{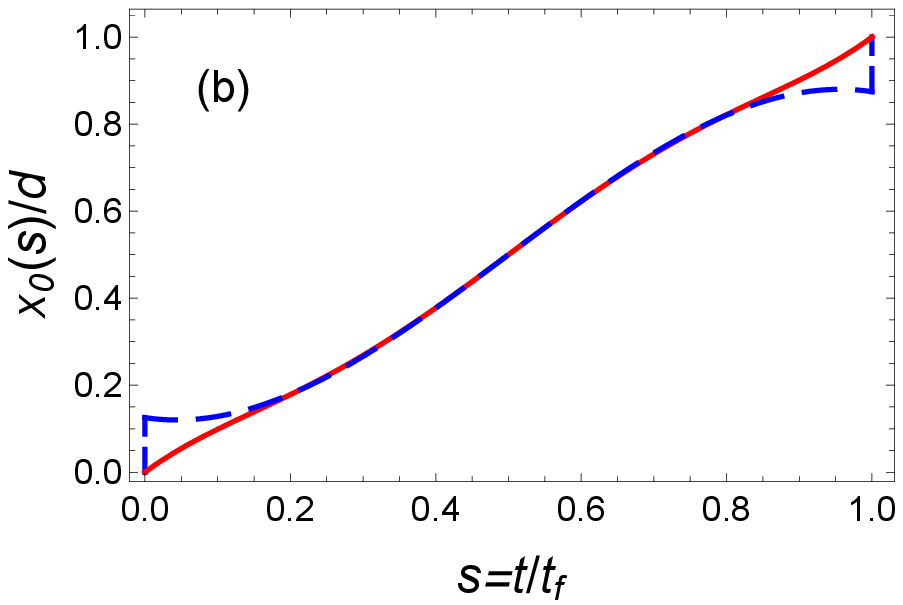}
		\caption{\label{figc1} Fast transport of atoms in a moving harmonic trap: Comparison of the trajectories of (a) the center of mass $x(t/t_f)/d$ and (b) the trap center $x_0(t/t_f)/d$, obtained from the OCT formalism by minimizing the time-averaged potential energy (blue dashed line) and using the IE approach (red solid line) based on a fifth-order polynomial ansatz. Parameters: $\omega_0=2\pi \times 50$ Hz and $t_f=22$ ms.}
	\end{center}
\end{figure}

\begin{eqnarray}
\label{x0-unbounded}
x_0 (t) \!=\! \left\{\begin{array}{lll}
0, & t \leq 0
\\
\displaystyle \!\!\! \left( \!\! 1-\frac{2t}{t_f} \! \right) \! \frac{6d}{\omega_0^2 {t_f}^2} \!+\!\! \left(\!\!3 \!-\! \frac{2t}{t_f}\! \right) \!\! \frac{t^2 d}{{t_f}^2}, & 0\!<\!t\!<\!t_f
\\
d, & t \geq t_f.
\end{array}\right.
\end{eqnarray}
In Fig.~\ref{figc1}, we plot the OCT center of mass along with the bottom trap trajectories for some specific values using blue dashed lines. It is worth noticing that according to our optimal solution the trap center has to include two sudden jumps at initial and final time. With such an optimization performed for an unbounded control function, we get the following lowest time-averaged potential energy

\begin{equation}
\label{EnUnbound}
\overline{E_p}^{(OCT)}=\frac{6m d^2}{\omega_0^2 {t_f}^4}.
\end{equation}
In Fig.~\ref{figE}, we also plot this minimal time-averaged potential energy as a function of the final time $t_f$ as a blue dashed line.

\begin{figure}[t]
	\begin{center}
		\centering\includegraphics[width=0.45\textwidth]{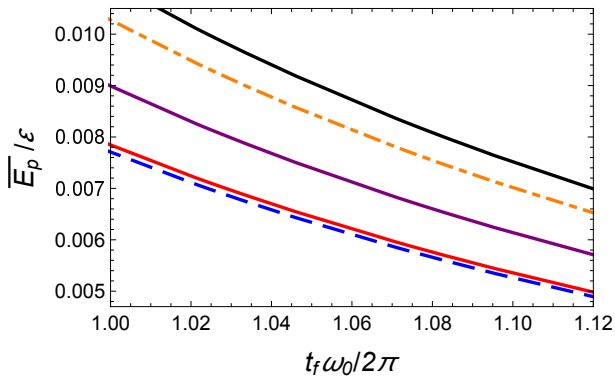}
		\caption{\label{figE} Fast transport of atoms in a moving harmonic trap: Time-averaged potential energy $\overline{E_p}/\varepsilon$ (normalized to $\varepsilon=m \omega_0^2 d^2/2$) as a function of final time $t_f$ by using different protocols: time-optimal (orange dash-dotted line), energy-minimization with unbounded constraint (blue dashed line), and IE approaches with a fifth-order polynomial (black upper solid line), a seventh-order polynomial (purple solid line), and nineteenth-order polynomial (red lower solid line). Same parameters as Fig.~\ref{figc1}.}
	\end{center}
\end{figure}


\subsection{Comparison between IE and OCT}

In this section, we use the freedom in the interpolation function that enters IE solutions to approach the solution of the optimal control theory associated to a minimization of the time-averaged potential energy.

\subsubsection{IE with polynomial ansatzs}

For this purpose, we first enlarge the parameter space of the polynomial ansatz that fulfills the boundary conditions (\ref{con0}) and search for the optimal values of the coefficients that minimize the time-averaged potential energy.

To satisfy the six boundary conditions (\ref{con0}) the minimal order of the polynomial interpolation function is five (see Eq.~(\ref{5orderpoly})). In Fig.~\ref{figc1}, we plot the center of mass, $x(t/t_f)/d$, and trap center, $x_0(t)$, trajectories as a function of time using red solid lines. The corresponding time-averaged potential energy is $\overline{E_p}^{(P5)}=1.42\overline{E_p}^{(OCT)}$
which is significantly larger than the minimal potential energy given by Eq.~(\ref{EnUnbound}). It is represented as a black solid line in Fig.~\ref{figE}.

In order to further reduce the time-averaged potential energy, we enlarge the parameter space, while keeping the six boundary conditions satisfied. We search for a solution of the forme  $x(t)=d [\sum\nolimits_{n=0}^{7} a_n (t/t_f)^n$]. By applying the boundary conditions (\ref{con0}), we have $a_0=a_1=a_2=0$, $a_5=21-6 a_3-3 a_4$, $a_6=-35+8 a_3+3 a_4$, and $a_7=15-3 a_3- a_4$. The time-averaged potential energy can be explicitly worked out:

\begin{figure}[t]
\begin{center}
\centering\includegraphics[width=0.4\textwidth]{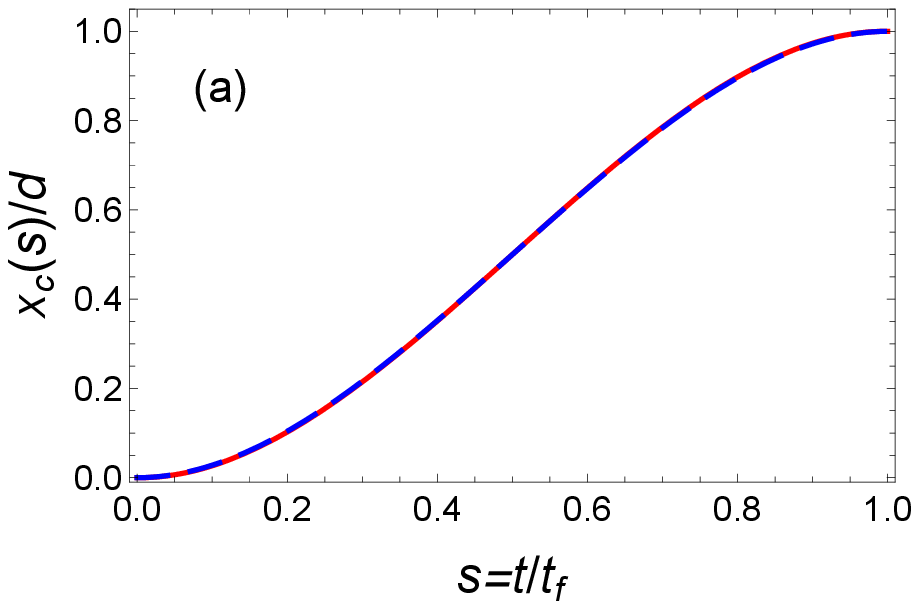}
\centering\includegraphics[width=0.4\textwidth]{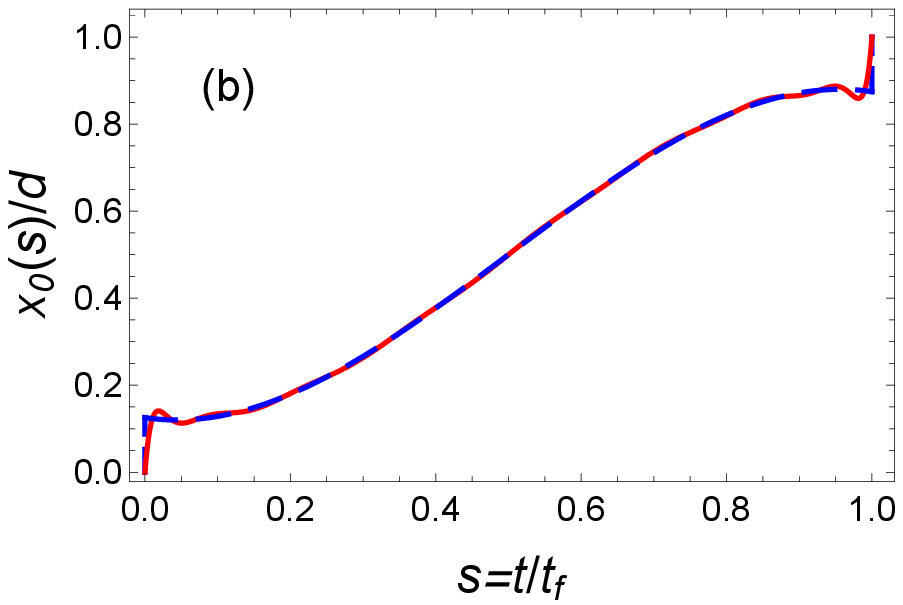}
\caption{\label{figc2} Fast transport of atoms in a moving harmonic trap: Comparison of trajectories of mass of center (a) and trap center (b), calculated from the OCT formalism (blue dashed line) and the IE approach (red solid line) with a 19th order polynomial ansatz. Same parameters as Fig.~\ref{figc1}.}
\end{center}
\end{figure}

\begin{eqnarray}
\nonumber
\overline{E_p}(a_3,a_4) &= &\left[ 7 +\frac{17}{77}(a_3-21)^2+\frac{4}{385}(a_4+70)^2 
\right.\\
&+& \left. \frac{(a_3-21)(a_4+70)}{11} \right] \frac{m d^2}{\omega_0^2 {t_f}^4}.
\end{eqnarray}
The minimization of this energy yields $a_3=21$ and $a_4=-70$ and $\overline{E_p}^{(P7)}\simeq 1.16\overline{E_p}^{(OCT)}$. This curve as a function of the final time is represented  in Fig.~\ref{figE} as a purple solid line. It provides a clear improvement with respect to the fifth-order polynomial solution. A priori, it is possible to further improve the optimization using a higher order polynomial ansatz. For instance, using a 19th order well-optimized polynomial, we have found $\overline{E_p}^{(P19)}\simeq 1.018\overline{E_p}^{(OCT)}$. In Fig.~\ref{figc2}, we have plotted  the corresponding time-dependent trajectories $x_c(t)$ and $x_0(t)$. However, the increase of the polynomial order requires a minimization with an increasing number of parameters. This is somehow cumbersome. In the following section, we propose another type of interpolating function inspired by the OCT solution and yielding astonishing results.

\subsubsection{IE with hyperbolic ansatz}

\begin{figure}[b]
\begin{center}
\centering\includegraphics[width=0.408\textwidth]{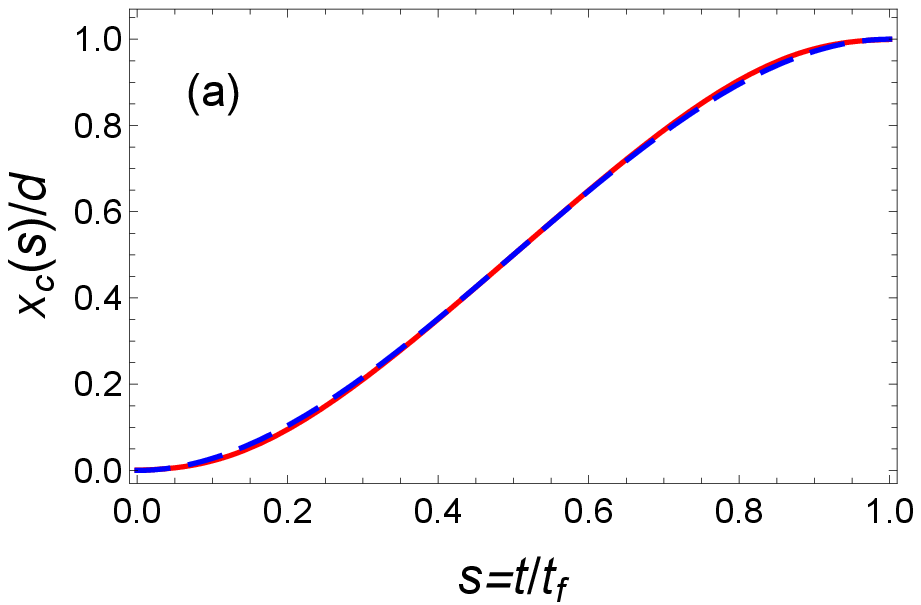}
\centering\includegraphics[width=0.4\textwidth]{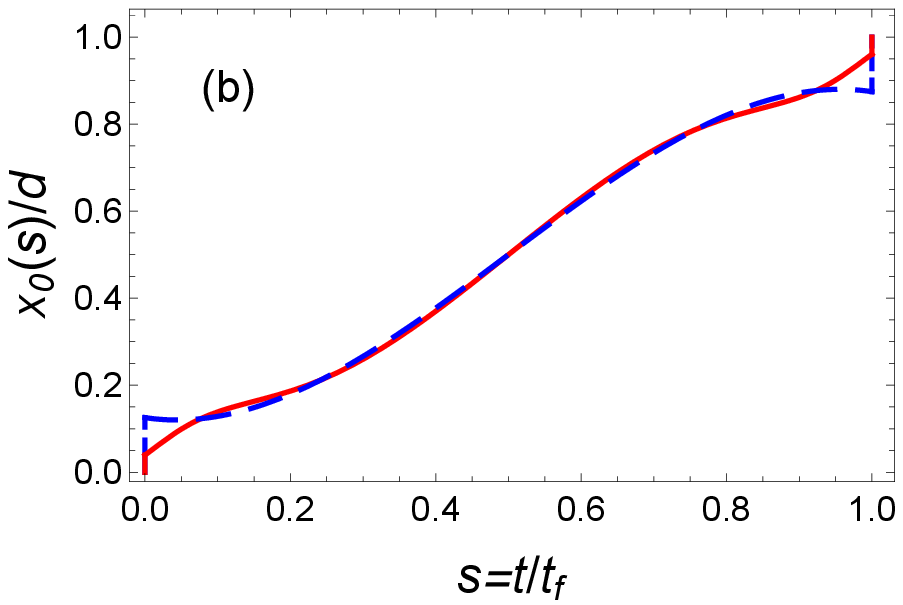}
\caption{\label{figct2} Fast transport of atoms in a moving harmonic trap: Comparison of trajectories of mass of center (a) and trap center (b), calculated from the OCT formalism (blue dashed line) and the IE approach with the optimized hyperbolic-function protocol in Eq.~ (\ref{xc-tanh}) (red solid line). The ``magic'' values are $a_1=1.2$, $a_2=1.25$,
	and the other parameters are the same as those in Fig. \ref{figc1}.}
\end{center}
\end{figure}

\begin{figure}[t]
	\begin{center}
		\centering\includegraphics[width=0.45\textwidth]{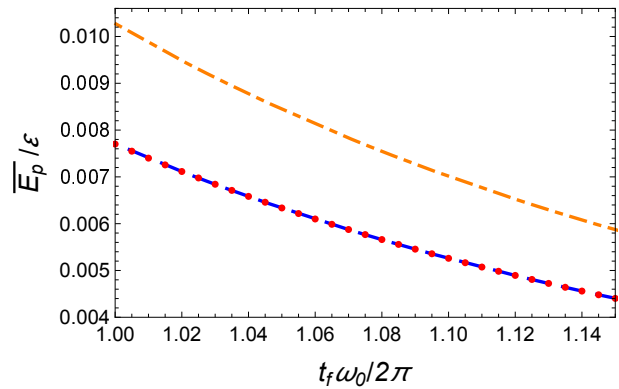}
		\caption{\label{figE2} Fast transport of atoms in a moving harmonic trap: Time-averaged potential energy $\overline{E_p}(t_f)/\varepsilon$ (normalized to $\varepsilon=m \omega_0^2 d^2/2$) calculated from different protocols: time-optimal  (orange dash-dotted line), energy-minimization with unbounded constraint (blue dashed line), and IE approach based on a hyperbolic-function-ansatz by choosing the ``magic'' values $a_1=1.2$ and $a_2=1.25$ (marked red point). Same parameters as Fig.~\ref{figE}.}
	\end{center}
\end{figure}

In this subsection, we apply the IE approach using the following hyperbolic-function

\begin{equation}
\label{xc-tanh}
x(t)=\frac{d}{2} \tanh \left\{a_1 \tan\left[\frac{\pi}{a_2} \left(\frac{t}{t_f}-\frac{1}{2}\right) \right]\right\}+\frac{d}{2},
\end{equation}
where $a_2>1$ to avoid any singularity. Interestingly, the choice of the parameter $a_2$ enables one to mimic a jump at initial and final time. This class of solution with the possibility of an initial and final offsest and with similar symmetry as the optimal function provides a very performant class of functions for the optimization. The freedom provided by the two parameters $a_1$ and $a_2$ enables one to reduce the time-averaged potential energy while satisfying the two boundary conditions $x(0)=0$ and $x(t_f)=d$. Such an optimization gives $a_1=1.2$ and $a_2=1.25$. The corresponding trajectories $x(t)$ and $x_0(t)$ are plotted in Fig.~\ref{figct2}, and 
the mean potential energy is represented in Fig.~\ref{figE2} with marked red points. This ansatz provides a solution that nearly coincides with the exact solution, $\overline{E_p}^{(hyp)}\simeq 1.0001\overline{E_p}^{(OCT)}$. 

For this transport problem, we have shown how the freedom on the interpolation ansatz enables one to optimize extra constraints such as the mean energy whilst fulfilling the boundary conditions. The choice of the ansatz has a strong impact. One could naively think that a very high order polynomial could always provide a succesful strategy. However, we have shown on this example that the convergence may be quite slow with the degree of the polynomial, and that the investigation of other shapes with a few adjustable parameters can easily outperform the polynomial interpolation for a give constraint. 

\section{Spin dynamics in the presence of dissipation}

In contrast with the previous sections, we address in the following an example dealing with the control of internal degrees of freedom. Optimal control provides a powerful tool to solve time-optimal and energy-optimal problems in quantum two-level and three-level systems \cite{JMP,Damme,Damme18,SpinStefanatos2009}. Such result can be directly extended to two uncoupled \cite{Damme18}  and coupled \cite{StefanatosPRA19} spins with similar approach. Using numerical optimal algorithm, robust optimal control can also be designed that accounts for inhomogeneous boarding and/or dissipation \cite{Owrutsky,Damme,Lin,ShersonNJP}. Inverse engineering techniques have also been used for the fast and robust control of single spin  \cite{QiScirep} and two-interacting spins \cite{Xiaotong,QiScirep} in the presence of dissipation \cite{impens}. Systematic error or perturbation induced from the parameter fluctuatiosn, dephasing noise, bit flip can be further suppressed using  IE and OCT in atomic population transfer \cite{PRL2013D,njp2012,inverse13} and spin flip \cite{Xiaotong}.

\begin{figure}[t]
\centering
\centering\includegraphics[width=0.45\textwidth]{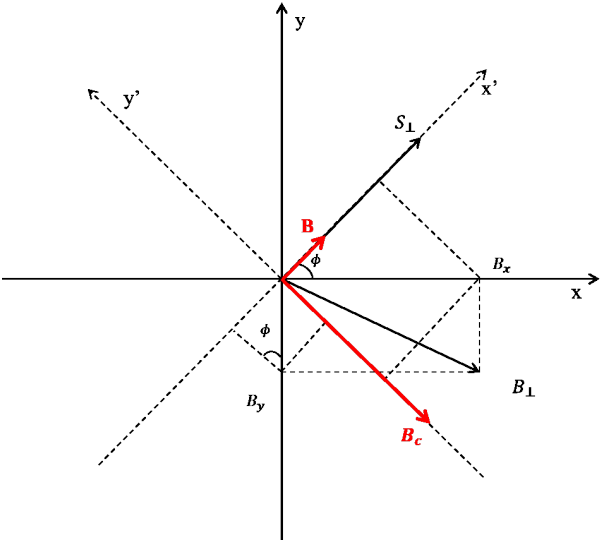}
\caption{Spin dynamics in the presence of dissipation: Equivalent magnetic field ($B, B_c$) of transverse magnetic field $B_{\perp}$.}
\label{figbt}
\end{figure}

Strictly speaking, the presence of dissipation rules out the possibility of an adiabatic evolution. However, the inverse engineering can still be applied.
In the following, we consider the control of a spin 1/2 (${\bf S}=(S_x, S_y, S_z)$through the appropriate design of the time-varying magnetic field components (${\bf B}=(B_x, B_y, B_z)$) for the desired boundary conditions. More precisely, we address the dissipative evolution of this spin in the presence of a strong transverse relaxation rate, $R>0$. As is commonly the case in NMR, the longitudinal relaxation rate is supposed to be negligible compared to the transverse one, and is here neglected \cite{SpinStefanatos2009}. Under those assumptions, the spin components obey the Bloch equations:
\begin{eqnarray}
\dot{S_x} &=& -R S_x-B_y S_z,\\
\dot{S_y} &=& -R S_y+B_x S_z,\\
\dot{S_z} &=& B_y S_x-B_x S_y.
\end{eqnarray}
Following Ref.~\cite{SpinStefanatos2009}, we recast the Bloch equations using spherical coordinates. 
For this purpose, we introduce the angles $\theta(t)$ and $\phi(t)$ such that ${\bf S}=(r\sin \theta \cos\varphi, r \sin \theta \sin \varphi, r \cos \theta)$ where $r$ denotes de length of the spin $r=\sqrt{S_x^2+S_y^2+S_z^2}$. It is convenient to decompose the transverse magnetic field $B_{\perp}=(B_x, B_y)$ into $B_{\perp}=(B, B_c)$, satisfying $B\parallel S_{\perp}$ and $B_c\perp S_{\perp}$ (see Fig.~\ref{figbt}): $B=(B_x/R)\cos \phi-(B_y/R)\sin \phi$ and $B_c=(B_x/R)\sin \phi+(B_y/R)\cos \phi$.
The Bloch equations can be readily rewritten with the variables $a=\ln r$, $\tan \theta=\sqrt{S_x^2+S_y^2}/S_z$, $\tan \phi=S_y/S_x$, and the normalized time $t=R t'$:

\begin{eqnarray}
\label{sysa}
\dot a &=& -[\sin \theta(t)]^2,\\
\label{systheta}
\dot \theta &=& B-\sin \theta(t) \cos \theta(t),\\
\dot \phi &=& B_c \cot\theta(t).
\end{eqnarray}
To ensure a spin rotation from an initial spin-up state to a given final target state, we shall use the boundary conditions

\begin{eqnarray}
\nonumber
\theta(0) &=& 0,\;\;a(0)=0,\;\;\theta(t_f)=\theta_f,\;\;\mbox{and}\\
a(t_f) &=& a_f=-\int^{t_f}_0[\sin\theta(t)]^2dt.
\label{theta}
\end{eqnarray}
It is worth emphasizing the fact that choosing the final spin length $a_f$ and orientation $\theta_f$ for a given final time $t_f$ may have no solution for finite resources. Indeed, if the driving by the magnetic field is not sufficiently strong, the dissipation will set an upper limit on the final spin length.

The field component $B_c$ is always perpendicular to $r_{\perp}$ and therefore only affects the spin rotation about the $z$-axis. The angle $\theta$ is responsible for the partial or total spin flip. To minimize the energy cost, the trajectory length shall be minimal. This latter condition sets the value of $B_c$ to zero which means $\phi=constant$. Basically, the IE technique amounts here to fixing the $\theta(t)$ function in accordance with the boundary conditions (\ref{theta}), and inferring the external magnetic field $B(t)$ from Eq. (\ref{systheta}).


\subsection{Energy minimization by OCT}

We consider here a given spin manipulation from $(a(0)=0, \theta(0)=0)$ to $(a_f, \theta_f)$ with the minimum magnetic field amplitude. For this purpose, we aim at minimizing the cost function

\begin{equation}
\label{energy}
E=\int^{t_f}_0 \frac{B(t)^2}{2}dt.
\end{equation}
Let's first recast this problem as a control problem involving a set of coupled first order equations. 
By defining the state variables $x_1=a$, $x_2=\theta$, and the control function
$u(t)=B(t)$, the system equations (\ref{sysa}) and (\ref{systheta}) is of the form $\dot{\textbf{x}} = \textbf{f}(\textbf{x}(t), u)$:
\begin{eqnarray}
\label{system-1spin}
\dot{x}_1  &=&  -\sin ^2 x_2,
\\
\label{system-2spin}
\dot{x}_2 &=& u- \sin x_2 \cos x_2,
\end{eqnarray}
and the cost function is
\begin{equation}
\label{energy-spin}
J=\int^{t_f}_0 \frac{u(t)^2}{2}dt.
\end{equation}
The corresponding Pontryagin Hamiltonian reads
\begin{equation}
\label{controlH-spin}
H_c=-\frac{1}{2}u^2- p_1 \sin^2 x_2+p_2(u-\sin x_2 \cos x_2),
\end{equation}
where $p_1$ and $p_2$ are the Lagrange multipliers fulfilling $\dot{\textbf{p}} = - \partial H_c/\partial \textbf{x}$ i.e. $\dot{p}_1 = 0$, $\dot{p}_2 = p_1 \sin (2x_2) +p_2 \cos (2 x_2)$. The maximum Pontryagin principle states for an unbounded control $u$ that $\partial H_c/\partial u=0$, i.e. $u=p_2$. In the absence of terminal cost, the optimal solution for this optimization between fixed initial and final states but without fixing the final time gives the extra condition $H_c [\textbf{p}(t),\textbf{x}(t),u(t)] =0$:
\begin{equation}
\label{optimalU}
p_2=(\sqrt{2p_1+\cos^2 x_2}+\cos x_2)\sin x_2.
\end{equation}

From Eq.~(\ref{system-2spin}), we deduce
\begin{equation}
\label{dotx2}
\dot{x}_2 = \sin x_2 \sqrt{\cos^2 x_2+2 p_1}.
\end{equation}
By combining Eqs.~(\ref{dotx2}) and Eq.~(\ref{system-1spin}), we find $d x_1=-\sin x_2/\sqrt{2p_1+\cos^2 x_2}d x_2$. After integration, this relation gives
\begin{equation}
r(\theta)=\frac{\cos\theta+\sqrt{2p_1+\cos^2\theta}}{1+\sqrt{2p_1+1}}.
\end{equation}
The (constant) value of $p_1$ is deduced self-consistently with the boundary conditions. The final time provided by OCT for an arbitrary target $r_f$ is determined by
\begin{equation}
\label{optimaltime}
t_f=\int^{t_f}_0 dt=\int^{\theta_f}_0\frac{1}{\dot\theta(\theta)} d\theta.
\end{equation}
{We note that the dissipation has an influence on the final time.}

\subsection{Case I: reaching the horizontal plane of the Bloch sphere}

In this subsection, we consider the transfer of the spin from the quantization axis to the horizontal plane. The boundary conditions are thus $\theta(0)=0$, $r(0)=1$ and $\theta(t_f)=\theta_f=\pi/2$. This choice sets the value of the constant $p_1$: $p_1^{\pi/2}=2{r_f}^2/(1-r_f^2)^2$.
To address a specific example, we consider in the following the final value $r(t_f)=r_f=e^{-2}$. The final time obtained from Eq.~(\ref{optimaltime}) suffers from a logarithmic divergence. To cure this problem, we shift the initial and final time by a small quantity $\varepsilon\ll 1$: $\theta(0)=\epsilon$ and $\theta_f=\pi-\epsilon$:
\begin{equation}
t_f^{\pi/2}=\frac{1-r_f^2}{1+r_f^2}\left[\ln\left(\frac{1+r_f^2}{r_f}\right)-\ln\epsilon\right]=8.60481849
\end{equation} 
for $\varepsilon=10^{-3}$. For this specific example, the cost function associated to this optimal solution (see Eq.~(\ref{energy-spin})) is
\begin{equation}
{E}^{(OCT)}_{\pi/2}=\frac{1}{1-{r_f}^2}=1.01866.
\end{equation}

\begin{figure}[t]
\begin{center}
\centering\includegraphics[width=0.4\textwidth]{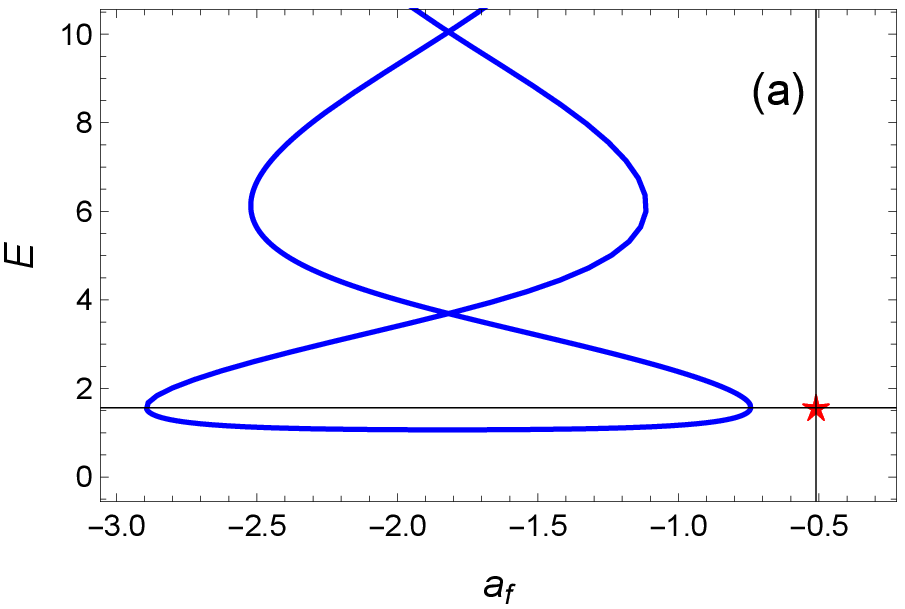}
\centering\includegraphics[width=0.4\textwidth]{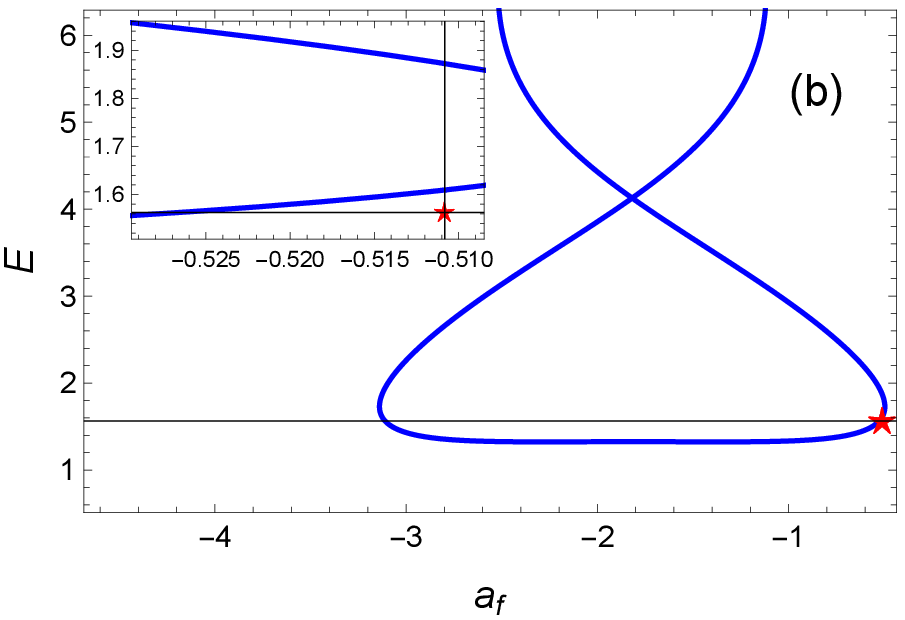}
\caption{\label{fig2} Spin dynamics in the presence of dissipation: Energy as a function of $a_f=\ln r_f$ for the same target state $(\theta_f, r_f, t_f)=(\pi/2, 0.6, 3.6357955)$. We compare the results obtained from the optimal control theory (red star) with the inverse engineering results involving two different polynomial ansatz fulfilling the boundary conditions. The energy curve are plotted for different values of the polynomial coefficient $a_1$: (a) for a second order polynomial ansatz and (b) for a third order polynomial ansatz with $a_3=0.1$. The inset in (b) shows the proximity of the inverse engineering result with that of the optimal control theory.}
\end{center}
\end{figure}

For comparison with the inverse engineering method, we propose, for the very same $t_f$, the following second order polynomial ansatz:

\begin{equation}
\label{poly1}
\theta(t)=a_1 t-\frac{a_1 t_f-\theta_f}{{t_f}^2} t^2.
\end{equation}
This ansatz fulfills the boundary conditions and has a single free parameter. The corresponding cost function, $E^{(P2)}$, is minimal for $a_1=-0.119582$: $E^{(P2)}=1.055E^{(OCT)}_{\pi/2}$.

\begin{figure}[t]
\begin{center}
\centering\includegraphics[width=0.41\textwidth]{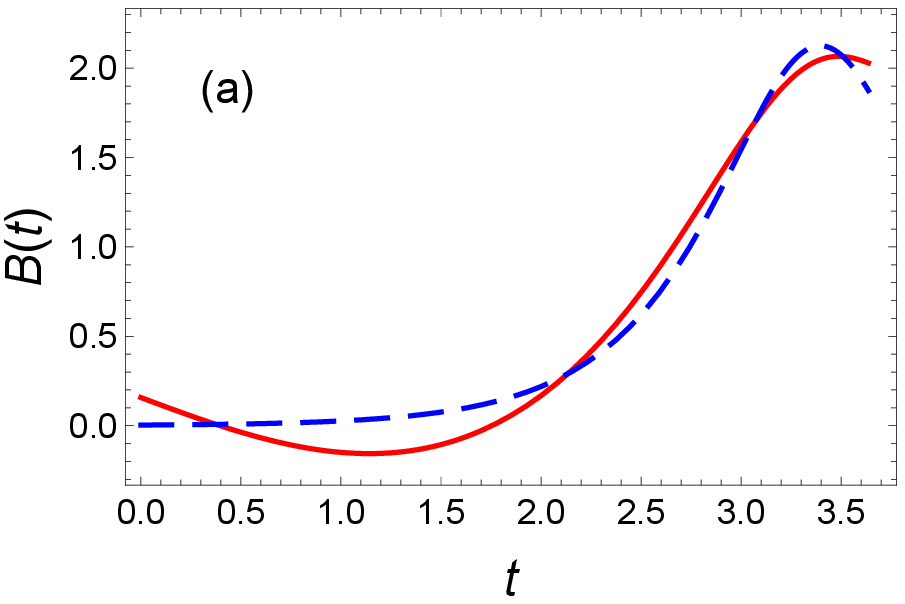}
\centering\includegraphics[width=0.41\textwidth]{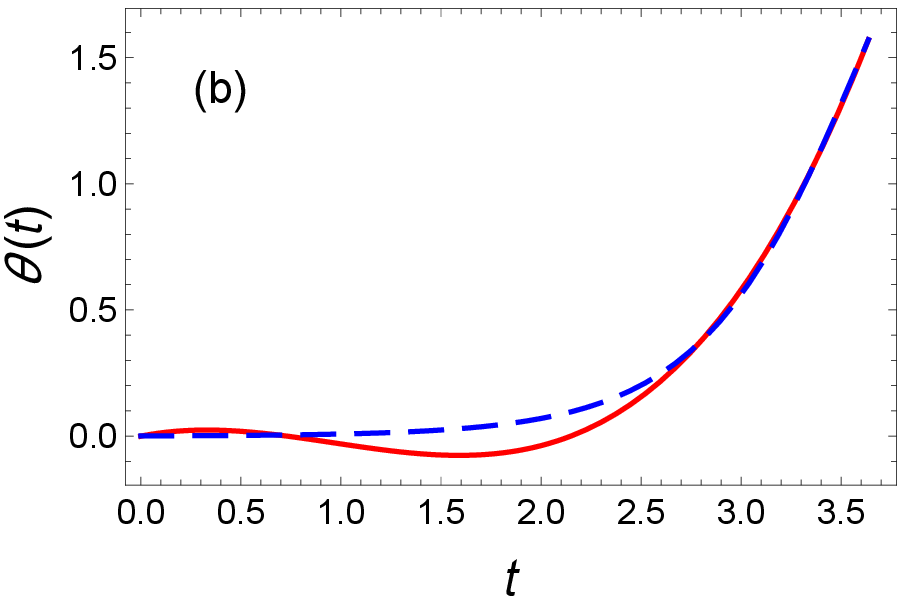}
\caption{\label{figH} Spin dynamics in the presence of dissipation: For a $\pi/2$ rotation, we plot (a) the magnetic field $B(t)$ and (b) the corresponding variable $\theta(t)$ obtained from an inverse engineering technique based on an optimized third-order polynomial (red solid line) and from the optimal control theory formalism associated to a mean energy minimization (blue dashed line).}
\end{center}
\end{figure}

\begin{figure}[t]
\begin{center}
\centering\includegraphics[width=0.45\textwidth]{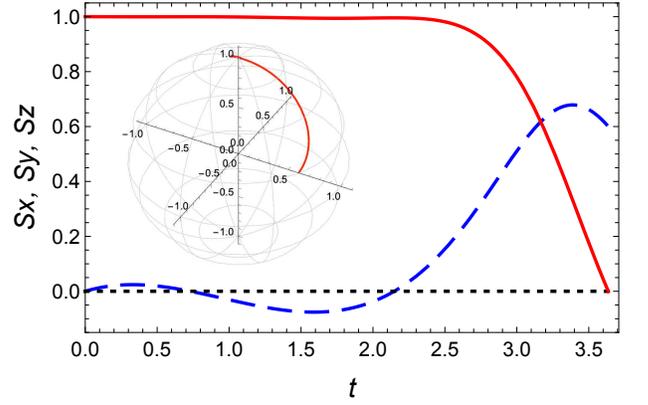}
\caption{Spin dynamics in the presence of dissipation: Time evolution of the spin components $S_z(t)$ (red solid line), $S_x(t)$ (blue dashed line), and $S_y(t)$ (black dotted line) under the magnetic field obtained from the inverse engineering method. Same parameters as Fig.~\ref{figH}. The inset depicts the corresponding spin trajectory on the Bloch sphere.}
\label{figHS}
\end{center}
\end{figure}

However, our simple polynomial ansatz provides an upper bound on the reachable values of $r_f$. This point is illustrated in Fig.~\ref{fig2} (a) where we plot the energy as a function of the logarithm of the final radius $a_f$ for different values of the free parameter $a_1$. For this example, the reachable range of values for $r_f$ is $[0.055;0.476]$. As a result, a target such as $r_f=0.6$ turns out to be out of reach. It is worth noticing that this limit is intimately related to the choice of the ansatz. For instance, we can choose a third-order polynomial ansatz:

\begin{equation}
\label{poly2}
\theta(t)=a_0+a_1 t+a_2 t^2+a_3 t^3,
\end{equation}
where $t_f$ is determined as previously ($t_f= 3.6357955$ for $r_f=0.6$) and, the coefficients $a_0=0$ and $a_2=-(a_1 t_f+a_3 {t_f}^3-\theta_f)/{t_f}^2$ are dictated by the boundary conditions (\ref{theta}). The extra freedom provided by the $a_3$ coefficient enables one to (1) reach the target and (2) minimize the cost function. With the values $a_3=0.1$ and $a_1= 0.15713222$, the cost function, $E^{(P3)}$, is quite close to the optimal value: $E^{(P3)}=1.03E^{(OCT)}_{\pi/2}$. In Fig.~\ref{fig2} (b), we plot the energy as a function of the free parameter $a_1$ for $a_3=0.1$. This curve defines a new interval of reachable $r_f$: $[0.043;0.608]$. The variable $\theta(t)$ and its corresponding magnetic field $B(t)$ obtained from the latter IE method are depicted in Fig.~\ref{figH}, and the associated spin trajectory on the Bloch sphere along with the spin components in Fig.~\ref{figHS}. Our results can be a priori further improved using an optimization on an even larger order polynomial.

\subsection{Case II: spin flip}

\begin{figure}[t]
\begin{center}
\centering\includegraphics[width=0.4\textwidth]{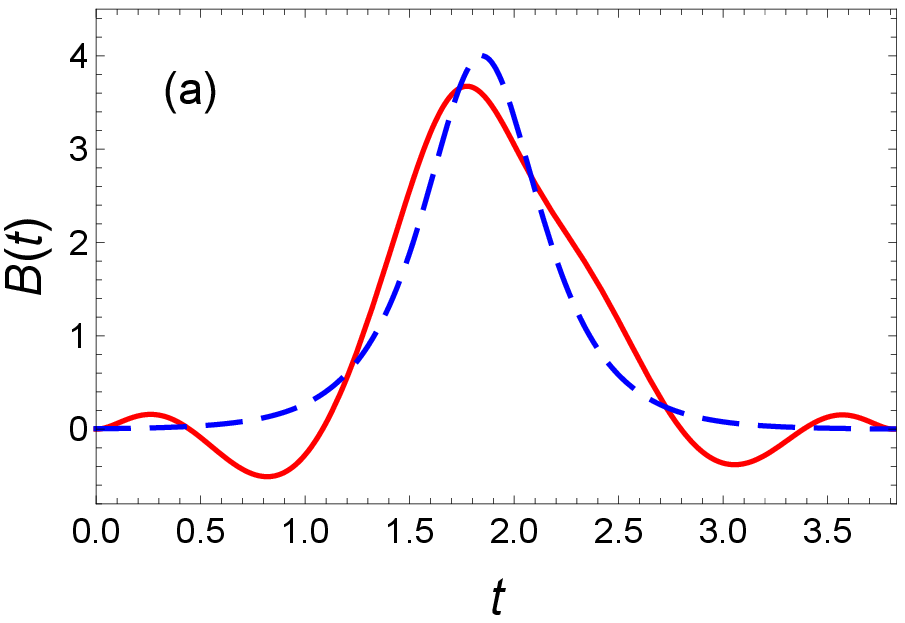}
\centering\includegraphics[width=0.4\textwidth]{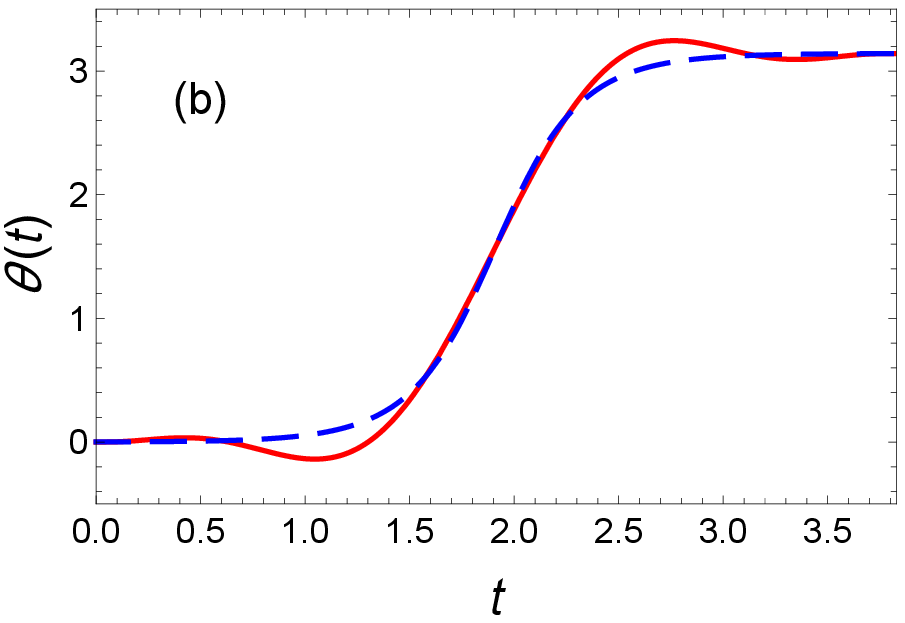}
\caption{\label{figflip2} Spin dynamics in the presence of dissipation: (a) The magnetic field $B(t)$ and (b) the variable $\theta(t)$ as a function of time for a minimal-energy spin flip. An optimal ninth-order polynomial has been used for $\theta(t)$ to apply the inverse engineering method (red solid line). The optimal solution is plotted as a blue dashed line.}
\end{center}
\end{figure}

\begin{figure}[t]
\begin{center}
\centering\includegraphics[width=0.4\textwidth]{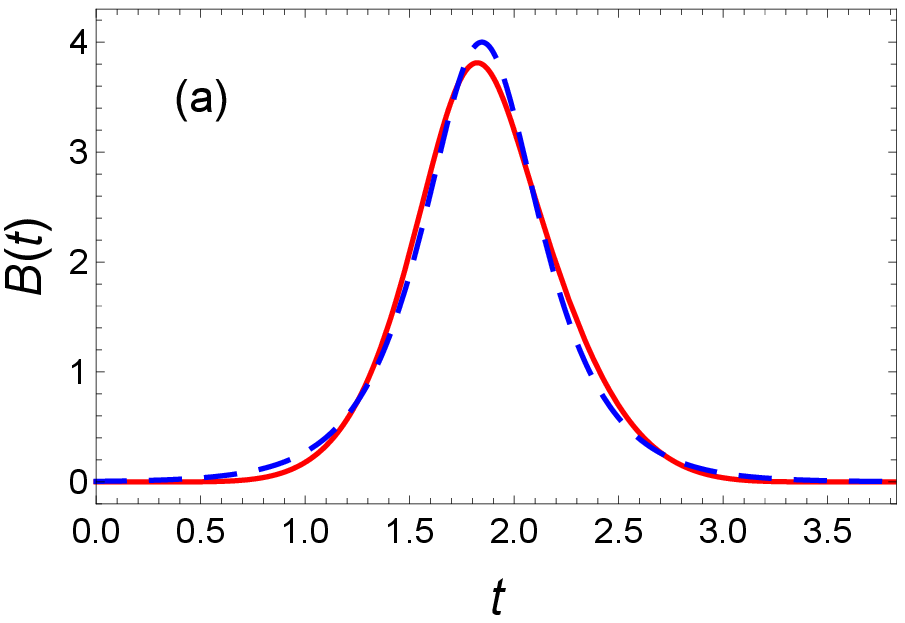}
\centering\includegraphics[width=0.4\textwidth]{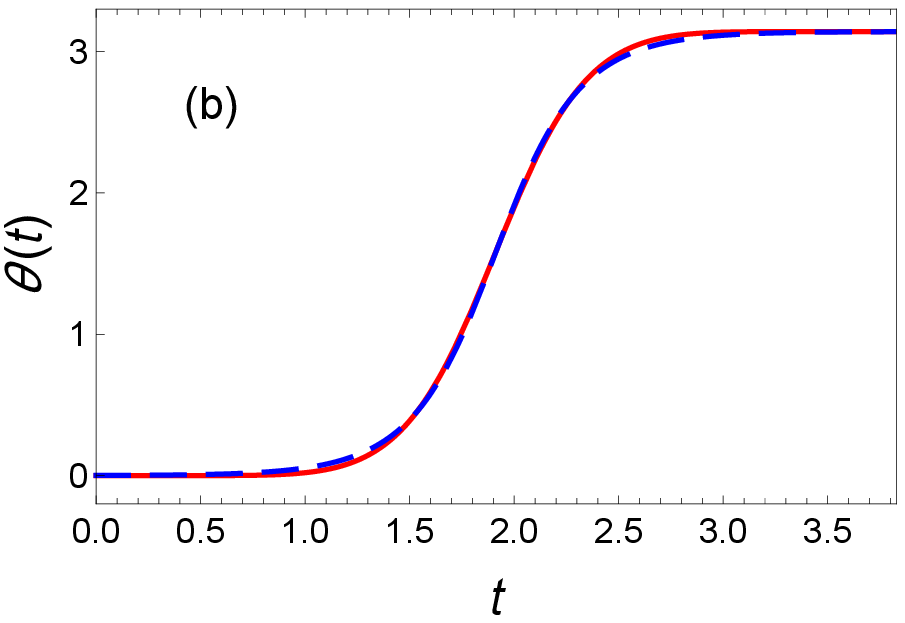}
\caption{\label{figflip3} Spin dynamics in the presence of dissipation: In the case of spin flip (b) obtained with magnetic field (a), compared with OCT (blue dashed line), an~$\tanh$ ansatz (instead of a polynomial) used in IE  approach (red solid line) is chosen to reduce  energy to $E=4.028$, with~``magic'' values $a_5=1.1$ and $a_1=3.104678$.}
\end{center}
\end{figure}

In this section, we consider a spin flip ($\theta_f=\pi$) for which the constant $p_1$ parameter is $p_1^{\pi}=2r_f/(1-r_f)^2$. With the same notations as previously, the final time reads (we use $r_f=0.6$ in the following)
\begin{equation}
t_f^{\pi}=\frac{1-r_f}{1+r_f}\left[\ln\left(\frac{(1+r_f)^2}{r_f}\right)-2\ln\epsilon\right]=3.8165858.
\end{equation}
The cost function associated to the optimal solution (see Eq.~(\ref{energy-spin})) is
\begin{equation}
{E}^{(OCT)}_{\pi}=\frac{1+r_f}{1-r_f}=4.0.
\end{equation}
This optimal solution is plotted as a blue line in {Fig.~\ref{figflip2}}. The optimal solution exhibits a smooth variations of $\theta(t)$ at initial and final  and a symmetry about $t_f/2$. This suggest to add the following extra condition to the polynomial ansatz for $\theta(t)$ for the inverse engineered solution:
\begin{eqnarray}
\nonumber
\theta(0) &=& 0,\;\;\theta(t_f/2)=\pi/2,\;\;\theta(t_f)=\pi,
\\
\label{boundaryflip}
\dot\theta(0)  &=& \dot{\theta}(t_f)=0,\;\;\mbox{and}\;\; \ddot{\theta}(0)=\ddot{\theta}(t_f)=0.
\end{eqnarray}
We have used a ninth-order polynomial to accommodate for the 7 boundary conditions listed above, an extra parameter is fixed by the final target radius, $r_f$. The remaining two free parameters are used to  minimize the energy. Knowing $\theta(t)$, we infer the magnetic field to be applied to drive the spin in accordance with our boundary conditions. As explicitly shown in Fig.~\ref{figflip2}, we find a bell shape for the magnetic field $B(t)$ associated to this $\theta(t)$. However, the curves remain relatively far from the optimal result. We find $E^{(P9)}=1.13E^{(OCT)}$. The ripples in the polynomial ansatz increase the energy and are difficult to remove by increasing the polynomial order. The convergence towards the optimal solution is therefore once again slow with the polynomial order. 

Alternatively, the shape obtained from OCT suggests that the following ansatz could be worth trying:
\begin{equation}
\label{tanh}
\theta(t)=\frac{\pi}{2} \tanh \left\{a_1 \tan\left[\frac{\pi}{a_5 t_f}(t-\frac{t_f}{2})\right]\right\}+\frac{\pi}{2}.
\end{equation}
Minimizing the energy, we find $E=1.007E^{(OCT)}$ with $a_5=1.1$ and $a_1=3.104678$. The comparison of this solution with its optimal counterpart confirms the proximity between the two approaches (see Fig.~\ref{figflip3}). 

\section{Conclusion}

In summary, we have investigated different implementations of the inverse engineering method and compare them with solutions deduced from the OCT for a given cost function. We have addressed in this manner the fast atomic cooling in harmonic trap, the atomic transport with a moving harmonic trap, and the spin control in the presence of dissipation. We have shown how the freedom on the ansatz inherent to inverse engineering techniques provide enough tunability to minimize a cost function while fulfilling the boundary conditions. We have systematically found class of functions with few adjustable parameters approaching the optimal control result with a relative excess of energy below one percent. Inverse engineered solutions are usually search as continuous and analytical functions which is a priori an asset for their practical use. However, we have also exhibit the possibility to design inverse engineered trajectories having initial and final jump to mimic the optimal control solution yielding solutions that are nearly undistinguishable from their optimal counterpart.



\acknowledgments{This research was funded by NSFC (12075145), STCSM (2019SHZDZX01-ZX04, 18010500400 and 18ZR1415500), Program for Eastern Scholar, Spanish Government PGC2018-095113-B-I00 (MCIU/AEI/FEDER, UE), Basque Government IT986-16. 
X. C. acknowledges Ram\'on y Cajal program (RYC-2017-22482).}


\end{document}